# Message Detection and Extraction of Chaotic Optical Communication Using Time-Frequency Analysis

Qingchun Zhao, Hongxi Yin[*]
School of Information and Communication Engineering
Dalian University of Technology
Dalian, P. R. China (hxyin@dlut.edu.cn)

*Abstract*—The security of chaotic optical communication using time-frequency (TF) representation is analyzed in this paper. The mean scalogram ratio (MSR) of TF representation and peak sidelobe level of MSR are defined to detect message. Algorithm for message detection and extraction is presented in detail. Two typical message encryption schemes, chaos masking and chaos modulation, are analyzed. The results reveal that it is not secure to transmit message when the message frequency locates at low power on power spectrum portrait. The proposed method is very useful for estimating the security level of message masking in chaotic optical communication.

## I. INTRODUCTION

Chaotic optical communication has attracted intensive research interest for its potential application in security encryption technologies based on the physical layer. This novel encryption technology has the following merits: broad bandwidth and large correlation dimension of chaotic carrier [1-3], compatibility with conventional optical fiber communication [4, 5], compact and integrated devices [6, 7], etc. Chaotic optical communication based on synchronization was experimentally achieved by fiber lasers and semiconductor lasers, separately [8, 9]. Giga Hertz sinusoidal messages encrypted by different schemes were presented in [10-13]. Recently, Argyris et al. successfully implemented a real-world chaotic optical communication for 1 Gbits/s transmission rate in 120 km commercial fiber-optic channel [14]. Integrated devices for chaos-based optical communication have been designed and fabricated [6, 7].

The aim of chaotic optical communication is to achieve secure transmission of message. Hence security level of this encryption technology is of great importance. From the standpoint of modern cryptography, it is necessary to attack the cryptographic system using cryptanalysis methods in order to analyse the security. This is no exception to chaotic optical communication. Chaotic synchronization between the emitter laser and receiver one can be easily achieved under strong injection strength [15]. However, this can also be used by eavesdropper to extract the message [16]. Subcarrier modulation was applied to enhance the security [17]. The feedback phase of semiconductor laser was considered a secret key [18]. Hurst exponent is utilized to analyze the security of chaos shift keying (CSK) and chaos modulation (CMO) [19]. Nonlinear filtering was employed to detect messages encoded by CMO [20].

In this paper, an algorithm based on time-frequency (TF) representation is proposed to detect and extract the message in chaotic optical communication. The security of chaos masking (CMA) and CMO encryption schemes are compared. The message extraction results are also shown.

This paper is organized as follows. The message detection and extraction algorithm is presented in section II. Section III shows the properties of chaotic carrier generated by a semiconductor laser. Section IV gives the numerical results of message detection and extraction. Section V concludes this paper and points out some scenarios for future work.

## II. ALGORITHM

Time-frequency (TF) analysis is a powerful tool to detect time-variation and non-stationary signals. As an emerging signal processing method, TF analysis has drawn more attention recently. TF representation provides the joint distribution information of time and frequency domain, which can distinctly depict the relationship between the frequency of signal and time. More information about TF analysis can be seen in [21-26].

Here, we plot the TF representation using continuous wavelet transformation. In order to detect the message hidden in chaotic carrier, we define the mean scalogram ratio (MSR) and peak sidelobe level (PSL) to quantificationally depict the TF representation. The message detection and extraction algorithm is presented as follows:

**Step 1: Continuous wavelet transform**
The continuous wavelet transform of the transmitted signal, i.e., the optical power in the chaotic optical communication channel $P(t)$, is given by

This work was supported in part by the National Natural Science Foundation of China (NSFC) under Grant 60772001 and Open Fund of State Key Laboratory of Advanced Optical Communication Systems and Networks (Peking University) and Scientific Research Start-up Fund of Dalian University of Technology for Introduced Scholars, P. R. China.

$$\mathrm{WT}_P(a,b) = \langle P(t), \psi_{a,b}(t) \rangle$$
$$= \frac{1}{\sqrt{a}} \int_{-\infty}^{+\infty} P(t) \psi^*\left(\frac{t-b}{a}\right) dt, \qquad (1)$$

where $a$ and $b$ represent the scale parameter and time shifting, respectively. The mother wavelet function $\psi(t)$ is chosen to be complex Morlet wavelet, which has the properties of high resolution and low aliasing. This function has the following form:

$$\psi(t) = \frac{1}{\sqrt{\pi f_b}} e^{2i\pi f_c t} e^{-t^2/f_b}, \qquad (2)$$

where $f_b$ and $f_c$ are the bandwidth parameter and wavelet center frequency, respectively.

**Step 2: TF representation**

The scalogram (SC) of the time trace $P(t)$ is defined by the module of the coefficients of wavelet transform obtained by Step 1:

$$\mathrm{SC}_P(a,b) = |\mathrm{WT}_P(a,b)|. \qquad (3)$$

This SC is a method to design the TF representation, which can present the frequency of signal varying with time.

**Step 3: Mean scalogram ratio**

Conventional signal detection methods are based on observation of the TF representation of the collected time trace. This is because the power of the expected signal is not very weak compared with the background noise. Furthermore, the intension of chaotic communication is to transmit small amplitude message. Therefore, the observation of the TF representation of transmitted signal in chaotic optical communication to extract the hidden message is not very effective. Considering the characteristics of chaotic optical carrier on the TF representation, we define the MSR

$$\eta_i = \overline{\mathrm{SC}(f_i)} \bigg/ \sum_{m=i-2}^{i+2} \overline{\mathrm{SC}(f_m)}, \qquad (4)$$

where the numerator denotes the mean of scalogram corresponding to the $i^{\mathrm{th}}$ frequency on the TF representation, while the denominator denotes the sum of neighbor values of the numerator including itself. Then the function between MSR and frequency can be obtained. We can obtain the signal frequency by locating the peak of this function. However, when the signal amplitude is rather small, the peak is not very sharp compared with the sidelobe. Hence we introduce peak sidelobe level (PSL) to further quantify, as described by Step 4.

**Step 4: Peak sidelobe level**

The PSL of frequency-MSR function is given by PSL=$10\log_{10}(\eta_P/\eta_S)$, where $\eta_P$ and $\eta_s$ denote the peak of $\eta$ and sidelobe one, respectively. After many numerical simulations, we find that the PSL of MSR of chaotic carrier is nearly zero. However, the PSL increases after message hidden in chaos. Therefore, the PSL of the transmitted signal is lager than that of chaotic carrier. Message exists and its frequency can be obtained by locating the peak of MSR.

**Step 5: Message extraction**

Band pass filter (BPF) whose central frequency is set to the peak of MSR is used to directly extract message from the time trace. The signal-to-noise ratio (SNR) is used to describe the message extraction, which is defined as SNR=$10\log_{10}(P_S/P_N)$, where $P_S$ and $P_N$ denote the mean power of the signal and noise, respectively.

## III. CHAOS CARRIER GENERATION

The dynamical characteristics of the emitter are described by the well-known Lang-Kobayashi rate equations with feedback term:

$$\frac{dS(t)}{dt} = \frac{\beta\Gamma N(t)}{\tau_N} + \frac{\Gamma g(N(t)-N_0)}{1+\varepsilon S(t)} S(t) - \frac{S(t)}{\tau_p} + 2\frac{\kappa}{\tau_l}\sqrt{S(t)S(t-\tau)}\cos(\theta(t)), \qquad (5)$$

$$\frac{d\phi(t)}{dt} = \frac{\alpha}{2}\left[\frac{\Gamma g(N(t)-N_0)}{1+\varepsilon S(t)} - \frac{1}{\tau_p}\right] - \frac{\kappa}{\tau_l}\sqrt{\frac{S(t-\tau)}{S(t)}}\sin(\theta(t)), \qquad (6)$$

$$\frac{dN(t)}{dt} = \frac{I}{qV} - \frac{N(t)}{\tau_N} - \frac{g(N(t)-N_0)}{1+\varepsilon S(t)} S(t), \qquad (7)$$

$$\theta(t) = 2\pi\tau c/\lambda + \phi(t) - \phi(t-\tau), \qquad (8)$$

where $S(t)$ is the number of photon, $N(t)$ is the carrier density, and $\phi(t)$ is the phase. $\kappa=(1-r^2)r_f/r$ denotes the feedback strength, where $r$ and $r_f$ represent the amplitude reflectivity of the laser cavity and the amplitude reflectivity of the external reflector, respectively. The following parameters were employed in simulations: $r$=0.3, $r_f$=0.01, central wavelength $\lambda$=1550 nm, spontaneous emission factor $\beta$=10$^{-5}$, optical confinement factor $\Gamma$=0.4, feedback delay $\tau$=4 ns, transparency carrier density $N_0$=4×10$^5$ μm$^{-3}$, threshold current $I_{th}$=12 mA, differential gain $g$=2.125×10$^{-3}$ μm$^3$ns$^{-1}$, carrier lifetime $\tau_N$=2 ns, photon lifetime $\tau_p$=2 ps, round-trip time in laser intracavity $\tau_l$=9 ps, linewidth enhancement factor $\alpha$=5.5, gain saturation parameter $\varepsilon$=3×10$^{-5}$ μm$^3$, and active layer volume $V$=150 μm$^3$.

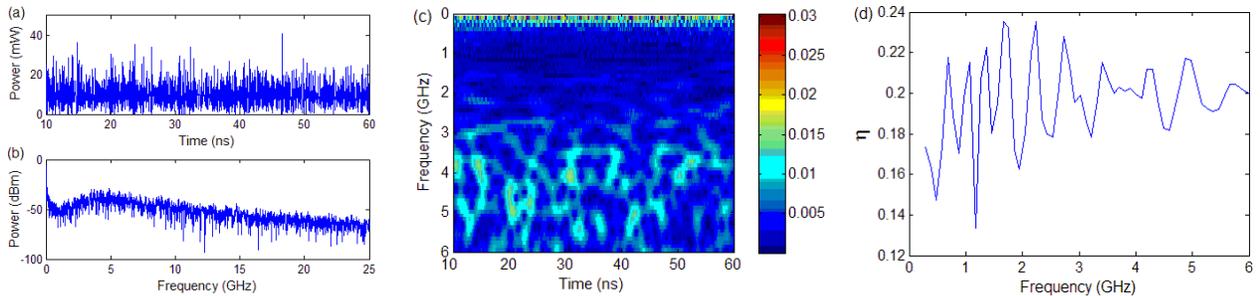

Figure 1. Chaotic carrier. (a) time trace, (b) power spectrum, (c) TF representation, (d) MSR.

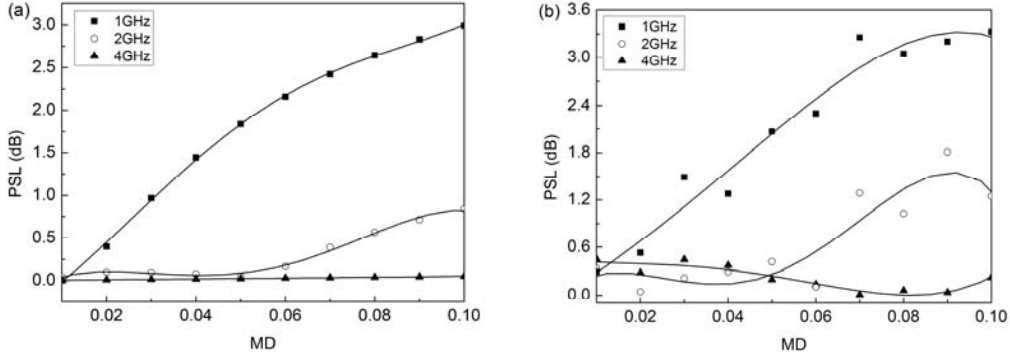

Figure 2. PSL as a function of modulation depth (MD) for different message frequency. (a) CMA, (b) CMO.

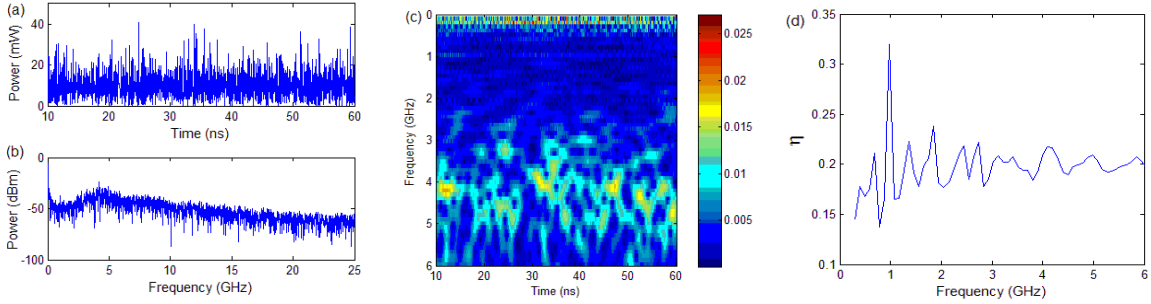

Figure 3. Detection of 1 GHz message encrypted by CMO. (a) time trace, (b) power spectra, (c) TF representation, (d) MSR.

Figure 1 exhibits the features of chaotic carrier generated by the emitter. As shown in Fig. 1 (a), the time trace of chaotic carrier is noise-like, which is expected to securely transmit small amplitude messages. The broadband spectrum indicates that ~ GHz message can be masked by the carrier. The TF representation of chaotic carrier is shown in Fig. 1 (c). For the frequency range of 0.5 GHz~2.5 GHz, the SC has no distinct difference, which is not suitable for message masking as demonstrated in Section Ⅳ. While for 2.5 GHz~5.5 GHz, some areas of big SC emerge. This frequency range with irregular SC can hide message effectively. The frequency more than 6 GHz is beyond the bandwidth of the carrier 5.93 GHz. The peak of MSR is not evident compared with the sidelobe, as shown in Fig. 1 (d). The PSL of the MSR is 0.0011 dB, which is used as the threshold of message detection.

## IV. MESSAGE DETECTION AND EXTRACTION

CMA and CMO are two typical encryption schemes for chaotic optical communication. After the synchronization between the twin lasers emitter and receiver, the message is added to the chaotic carrier by an electro-optical modulator if the message is an electrical signal or a beam splitter if the message is an optical signal [14]. This encryption is the so-called CMA described by $P(t)=P(t)(1+MD\sin(2\pi ft))$, where $P(t)$, MD and $f$ are the output power of the chaotic carrier, modulation depth, and frequency, respectively. While for CMO, the message is encrypted using a bias-tee by directly modulated the bias current of the emitter following this expression $I=I_b(1+MD\sin(2\pi ft))$, where $I_b$ is the bias current of the emitter [13].

We use the method introduced in Section II to quantify the security level of these two encryption schemes CMA and CMO. 1 GHz, 2 GHz and 4 GHz sinusoidal messages are chosen for the power of chaotic carrier at 1 GHz is low and at 4 GHz is high. We numerically study the PSL of MSR under different message amplitudes and frequencies. As can be seen form Fig. 2 (a), the PSL for 1 GHz message increases rapidly with the increasing MD. However, the PSL for 4 GHz message nearly keeps steady (nearly zero) during the whole variation of MD. As shown in Fig. 2 (b), the variations of PSLs are as disciplinary as CMA after data fitting. However, the variation of PSL for CMO is not very regular as CMA before fitting. This is caused by the difference between these two encryption schemes. To CMO, the modulation of current changes the dynamic characteristics of emitter especially when the MD is large. While the modulation of CMA is outside the emitter, i.e., the modulation has no effect on the emitter.

As an example, Fig. 3 shows the detection of 1 GHz message encrypted by CMO with MD = 0.04. There exists 1 GHz frequency from the TF representation. Due to the frequency resolution of wavelet transform, there is a width for

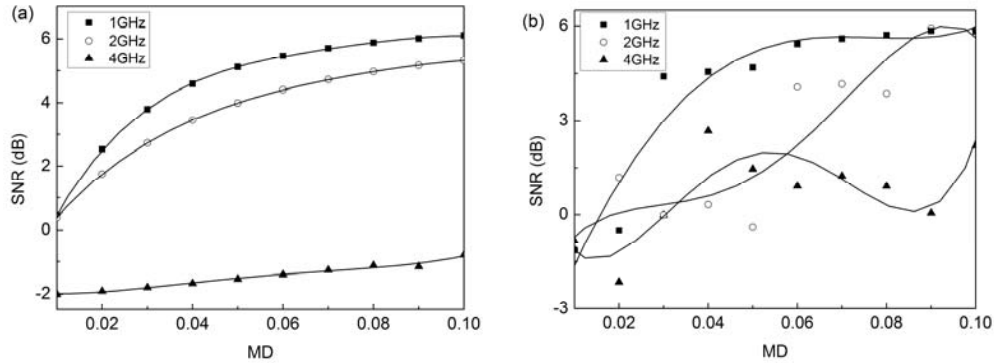

Figure 4. SNR of the extracted message using BPF versus modulation depth (MD) for different message frequency. (a) CMA, (b) CMO.

1 GHz. Hence the peak of the MSR locates at around 1 GHz. According to this, a band pass filter can be used to extract the hidden message.

In order to further demonstrate the effectiveness of the method proposed in this paper, we use band pass filter (BPF) to extract the message. The Chebyshev Type I digital filter is used to design the BPF. The central frequency of BPF equals to the peak of MSR. The other parameters for BPF keep constant during the variation of message amplitude and frequency. These parameters are band pass width 0.2 GHz, passband ripple 3 dB, and stopband attenuation 20 dB. Fig. 4 (a) shows the SNRs of the extracted message for CMA. The SNR for 1 GHz message is larger than that for other two frequency messages for certain MD. Note that all the SNRs for 4 GHz are below zero, which means that message extraction in practice is impossible. The message extraction for CMO is also affected by the message modulation of the injection current as shown in Fig. 4(b). Yet message extraction can also be realized when MD is larger than 0.03 for all three cases above.

## V. CONCLUSIONS

In summary, we propose a method based on TF representation to evaluate the security level of chaotic optical communication. We define the MSR of TF representation and PSL of MSR to detect the hidden message in chaotic carrier. Two typical encryption schemes CMA and CMO are analyzed using this method. The results indicate that it is not secure to transmit message when the message frequency locates at low power on power spectrum portrait, vice versa. Although the modulation of current affects the output of the emitter laser, message detection and extraction can also be achieved, especially when the peak for MSR are high.

The up-to-date chaotic optical communications are analog ones, i.e., the chaotic carriers are analog signals. To these communication systems, however, the receivers are not very sensitive to parameter mismatch except complete synchronization system. This insensitivity degrades the system security for Eve can also use an identical device with approximate parameters of the receiver to eavesdrop [16]. Yet digital communications are more practical and secure. The discretization of chaotic optical signal has been reported recently [27, 28]. The digital encryption based on discrete chaotic optical signal will enhance the security emphatically, which will become a hot topic in future.

ACKNOWLEDGMENT

Q. Zhao thanks Bin Ni at Shanghai Branch of National Instruments Co. Ltd. for useful discussions on time-frequency analysis.